\documentclass[10pt, a4paper]{article}

\usepackage{lrec-coling2024} 
\usepackage{multirow}

\title{Speech Corpus for Korean Children with Autism Spectrum Disorder: Towards Automatic Assessment Systems}

\name{Seonwoo Lee\textsuperscript{1}, Jihyun Mun\textsuperscript{1}, Sunhee Kim\textsuperscript{2}, Minhwa Chung\textsuperscript{1}} 

\address{\textsuperscript{1}Department of Linguistics, Seoul National University, \\ 
    \textsuperscript{2}Department of French Language Education, Seoul National University \\
    Gwanak-ro 1, Gwanak-gu, Seoul, Republic of Korea \\
    \{lsw5220, jhhh\_1202, sunhkim, mchung\}@snu.ac.kr\\}

\abstract{Despite the growing demand for digital therapeutics for children with Autism Spectrum Disorder (ASD), there is currently no speech corpus available for Korean children with ASD.
This paper introduces a speech corpus specifically designed for Korean children with ASD, aiming to advance speech technologies such as pronunciation and severity evaluation.
Speech recordings from speech and language evaluation sessions were transcribed, and annotated for articulatory and linguistic characteristics. Three speech and language pathologists rated these recordings for social communication severity (SCS) and pronunciation proficiency (PP) using a 3-point Likert scale. 
The total number of participants will be 300 for children with ASD and 50 for typically developing (TD) children.
The paper also analyzes acoustic and linguistic features extracted from speech data collected and completed for annotation from 73 children with ASD and 9 TD children to investigate the characteristics of children with ASD and identify significant features that correlate with the clinical scores.
The results reveal some speech and linguistic characteristics in children with ASD that differ from those in TD children or another subgroup of ASD categorized by clinical scores, demonstrating the potential for developing automatic assessment systems for SCS and PP.
 \\ \newline  }

\begin{document}

\maketitleabstract

\section{Introduction} 

Children with ASD (Autism spectrum disorder) exhibit deficits related to social communication skills \cite{socialcommunication}. 
In \citet{wetherby2006understanding}, several examples highlight challenges in initiating interactions with others, responding to social overtures, using appropriate vocal tones and gestures, understanding another person's perspective, and engaging in reciprocal conversations.
These deficits lead to fewer social interactions than typically developing (TD) children because children with ASD encounter greater difficulties when attempting to interact with others and display limited awareness of their social challenges. 

Speech production is one of the difficulties they have, which has received comparatively less attention \cite{wolk_phonological_2013} even though 80\% of children with ASD are able to produce at least one word for communication \cite{turner_follow-up_2006, charman_outcome_2005}.
They exhibit atypical prosody , includin the higher fundamental frequency (F0) and its greater variability \cite{diehl_acoustic_2012, fusaroli_is_2017, lyakso_perception_2017, bonneh_abnormal_2011, mccann_prosody_2003}; slower speech rate \cite{bone_spontaneous-speech_2012}, longer duration of utterances, frequent and prolonged pauses \cite{diehl_acoustic_2012, fusaroli_is_2017}; and poor voice quality \cite{bone_spontaneous-speech_2012}.
Articulatory and phonological skills are also delayed in that some children with ASD score below the lower limit of the normal range on standardized articulation tests \cite{rapin_subtypes_2009, cleland_phonetic_2010, shriberg_hypothesis_2011}. 
A substantial 41\% of school-aged children with ASD produce pronunciation errors that are not only developmentally delayed but also deviate from typical patterns \cite{cleland_phonetic_2010, wolk_phonological_2013}, which can significantly hinder effective communication.

The diagnosis of ASD reflects these aspects. 
According to \cite{american_psychiatric_association_diagnostic_2013}, ASD is diagnosed based on difficulties in communication and interaction, as well as restricted interests and repetitive behaviors, which affect daily functioning. 
In the clinical setting, standardized diagnostic tests, such as the Autism Diagnostic Observation Schedule, 2nd edition (ADOS-2) \cite{lord_autism_2012}, are conducted.
It not only measures the difficulties in social communication \cite{socialcommunication}, but also contains an item for prosody evaluation. 
However, evaluating children using standardized tests is challenging.
The shortage of expertise has led to delayed or missed diagnoses \cite{li2022automatic}.
Moreover, results of the standardized tests could be biased by subjectivity from caregivers or evaluators \cite{frigaux2019adiados}. 
The evaluation process lasting longer than an hour also increases children's and their caregivers' burden as well as decreases children's concentration.

Therefore, digital therapeutics (DTx) for children with ASD have been actively developed in the screening, diagnosis, and treatment of ASD \cite{dtx}. 
Recent studies for DTx systems for children with ASD have utilized a machine learning approach using video data.
\citet{kojovic2021using} introduced a machine learning approach that differentiated between children with ASD and TD by recognizing actions in video data.
In \citet{cilia2021computer}, visual data from eye-tracking was pivotal for the diagnosis of ASD. 
However, collecting video data costs a lot given that it necessitates specialized equipment like eye trackers or cameras, which can be also distracting for children \cite{albo2013comparing}. 
In contrast, speech data offers distinct advantages over video recordings.
Audio recordings are more cost-effective and straightforward than video recordings, as well as less intrusive \cite{clemente2008recording}.

In spite of the advantages of speech, there is an absence of a DTx system evaluating children with ASD based on speech data.
It can be attributed to the scarcity of speech corpora for machine learning.
There are several speech corpora for children with ASD. 
Dutch ASymmetries Corpus in the ASDBank \cite{kuijper_who_2015} is composed of audio recordings of 46 children with ASD. Each child was on a structured storytelling task. The basic demographics such as age and gender are available on the website.
The USC CARE corpus \cite{black_usc_2011} contains speech from 46 children with ASD and 14 children without ASD. The speech recordings were from entire ADOS sessions, which enabled the corpus to include ADOS code scores and final ADOS diagnosis.
The CSLU Autism speech corpus \cite{gale_improving_2019} is utilized for improving the performance of speech recognition. It contains 30 autistic children and 13 children with specific language impairments. The speech was recorded during a recalling sentence task, a sub-task of a standardized language assessment. As the task is short, the total duration of the speech is one and a half hours.

These corpora exhibit limitations in developing an automatic assessment system for the following reasons: 
First of all, the amount of speech from children with ASD is limited. While there are more than 40 children with ASD \cite{kuijper_who_2015, black_usc_2011}, the speech duration might be limited to improve the performance of machine learning models. For instance, the USC CARE corpus has a total duration of 50 hours, but this also includes speech from clinicians.
Another concern is the unavailability of clinical scores which is crucial for the purpose of machine learning, especially supervised learning. While some assessments have been implemented during the construction of the corpora, the clinical scores are not provided, except for the USC CARE corpus. However, the ADOS codes in this corpus were assessed by one of three psychologists, which may introduce some degree of subjectivity. 
Furthermore, there is an absence of clinical scores from speech and language pathologists, who possess specialized expertise in speech and language areas. No study analyzed the speech or language features of children with ASD in correlation with the clinical scores, accordingly.

As for the Korean language, any specialized speech corpus for children with ASD is not available yet, while Korean speech corpora for various disorders have been recently constructed including dysarthria \cite{qolt}, cleft palate \cite{cleft_palate}, and multiple speech disorders with various etiology, which is released by AI-Hub\footnote{Website: https://www.aihub.or.kr/}.

To overcome the absence of a suitable speech corpus for automatic assessment systems tailored for children with ASD, this paper presents a speech corpus composed of recordings of Korean children with ASD and clinical scores regarding social communication severity (SCS) and pronunciation proficiency (PP).
The speech within the corpus is then examined in terms of acoustic and linguistic attributes in relation to the clinical scores to figure out significant features for automatic assessment systems for children with ASD.
This corpus is the first speech corpus of children with ASD that includes perceptually evaluated SCS and PP by 3 speech and language pathologists (SLPs). 
Moreover, it is also the first speech corpus of Korean children with ASD.

\section{Speech Corpus of Children with ASD}
\subsection{Data Collection}
Aiming to capture speech and language traits of children with ASD, the speech recordings are obtained during speech and language evaluation sessions conducted by certificated SLPs at speech and language therapy centers in Korea.
Each session includes standardized tests for speech and language, which are widely used in Korea, measuring articulation, receptive and expressive language skills, and vocabulary understanding and production, as well as natural conversation in a clinical setting.
According to the child's chronological age and language development, different tests are conducted. Specific standardized tests are as follows:

\begin{itemize}
    \item Articulation: Assessment of Phonology and Articulation for Children (APAC) \cite{kim_assessment_2007}
    \item Language skills: Sequenced Language Scale for Infants (SELSI) \cite{selsi}, Preschool Receptive-Expressive Language Scale (PRES) \cite{pres}, Language Scale for School-aged Children (LSSC) \cite{lssc}
    \item Vocabulary: Korean version of Macarthur-Bates Communicative Development Inventories (K M-B CID) \cite{k_mb_cid}, Receptive \& expressive vocabulary test (REVT) \cite{revt}
\end{itemize}

The articulation evaluation is an essential part of each session to identify the distinctive pronunciation characteristics of children with ASD.
During the process, a child names pictures or repeats utterances provided by an SLP. 
Even if a child is unable to complete the articulation test, a portion of the test is still recorded.
A session generally lasts around 1 hour, and the duration of a child's utterances during a session is expected to be approximately 5 minutes.
However, the length of both the session and the child's speech can vary based on the child's chronological age and attention span.
Throughout the session, both children and SLPs have their speech recorded.
A Logitech Blue Yeti microphone is located at the ceiling's center, preventing a child from being distracted.
Starting in June 2022, the corpus will comprise 300 children with ASD and 50 TD children by the end of 2024, encompassing at least 25 hours of speech from children with ASD.
Children with ASD are determined based on either DSM-4 or DSM-5 criteria.
The corpus incorporates meta information including chronological age, gender, and language evaluation outcomes.

\subsection{Transcription and Annotation}
All the audio recordings of the sessions are transcribed and annotated by trained transcribers. 
Given the challenges of deciphering a child's intended speech, the transcription is done phonemically in the Korean alphabet. 
However, this approach poses limitations for analyzing linguistic aspects and pronunciation-related attributes of children, impeding the identification of inappropriate use of language or pronunciation errors. 
To overcome the limitation, a transition to orthographic transcription is being planned.
This shift will facilitate a more comprehensive analysis, enabling the extraction of linguistic traits for automatic SCS evaluation and automatic mispronunciation detection.

The transcriptions are enriched with annotations. 
Annotations for pre-processing the audio data include overlap, noise, low volume, as well as non-linguistic sounds like coughs or laughter. 
In order to probe linguistic attributes, immediate echolalia, off-topic utterances or delayed echolalia, exclamation, and long pauses are identified and labeled.
To identify speech sound errors that arise during the articulation evaluation, a target word and mispronounced phonemes are presented together, along with a special symbol.
Transcriptions and audio segments containing identifiable information are masked.

\subsection{Clinical Scores}
\subsubsection{Social Communication Severity Level}
The social communication severity of each child with ASD is evaluated by three SLPs holding a national certificate in speech and language pathology and possessing over three years of clinical experience, who did not participate in the data recording process. 
These evaluators classify the children's social communication severity based on their audio recordings and transcriptions.
The evaluation criteria are adopted from the social communication component of the diagnosis criterion of DSM-5 criteria \cite{american_psychiatric_association_diagnostic_2013}.
The severity is categorized into three levels, considering the extent of required support.
The levels are as follows: 
\begin{itemize}
    \item Level 1 (REQUIRING SUPPORT): Have difficulty initiating social interactions, may exhibit unusual or unsuccessful responses to social advances made by others, may seem to have decreased interest in social interactions.
    \item Level 2 (REQUIRING SUBSTANTIAL SUPPORT): Exhibit marked delays in verbal and non-verbal communication, have limited interest or ability to initiate social interactions, and have difficulty forming social relationships with others, even with support in place.
    \item Level 3 (REQUIRING VERY SUBSTANTIAL SUPPORT): Have very limited initiation of social interaction and minimal response to social overtures by others and may be extremely limited in verbal communication abilities.
\end{itemize}

The final severity levels are determined through a voting process, wherein a level is selected if it obtains agreement from more than 2 SLPs.

\subsubsection{Pronunciation Proficiency Score}
The three SLPs who conduct the SCS evaluation also assess pronunciation proficiency in children with ASD. 
The same audio recordings and transcriptions for SCS evaluation are used for PP evaluation.
The evaluation criteria are adopted from the item within ADOS-2 \cite{lord_autism_2012} on voice and intonation.
The criteria are outlined as follows:

\begin{itemize}
\item Score 1: The intonation is not peculiar or strange, displaying typical and appropriate modulation
\item Score 2: Less variation in pitch and tone modulation. Monotonic, exaggerated, or occasionally peculiar intonations are observed
\item Score 3: Unusual intonation or inappropriate pitch and stress result in markedly monotonous or mechanical sounds lacking inflection. The child emits peculiar cries or sounds.
\item Excluded: Limited voice production for the evaluation. While exhibiting normal crying, other vocalizations are scarce.
\end{itemize}
The final pronunciation scores are calculated by averaging all scores obtained by the three SLPs and subsequently rounding it.


\section{Initial Analysis of Acoustic and Linguistic Features}

\subsection{Current version of the corpus}
Starting from 2022 until now, 113 children with ASD (93 boys, 18 girls, and 2 not reported) and 9 children with TD (6 boys and 3 girls) have participated in audio recording.
The annotation and evaluation for clinical scores are complete for 73 children with ASD (59 boys, 12 girls, and 2 not reported) and all TD children.
Among children with ASD, 2 children's chronological age has not been reported yet.
Except for the 2 children, the mean chronological age of 73 children with ASD is 7;8 years.
For the 9 TD children, the mean chronological age is 7;10 years.
For children with ASD, the total number of utterances both from children with ASD and SLPs is 57,856 (58 hours and 22 minutes), and the number of utterances only from children with ASD is 20,841 (14 hours and 17 minutes). 
For children with TD, the number of utterances both from TD children and SLPs is 7,032 (6 hours 42 minutes) and that of utterances only from TD children is 2,231 (1 hour and 16 minutes).

The number of children in the TD group and the three ASD subgroups is summarized in Table \ref{group_info_table}.
For SCS, 73 children with ASD are categorized into three subgroups based on their SCS levels: level\_1 (L1), level\_2 (L2), and level\_3 (L3). 
A lower level indicates better social communication skills. 
During the analyses of PP, audio recordings of 21 children with ASD are excluded, who did not contain sufficient speech for evaluation. As a result, features from 52 children with ASR are used in the subsequent analysis. 
These children with ASD are further categorized into three subgroups based on their PP scores: score\_1 (S1), score\_2 (S2), and score\_3 (S3). 
A lower score indicates better pronunciation.

The inter-rater reliability of each clinical score is assessed using the Intraclass Correlation Coefficient (ICC). 
For SCS evaluation, the inter-rater reliability is 0.939 with a 95\% confidence interval of 0.910 to 0.960 (p$<$0.001).
Following the PP evaluation, the calculated ICC value for 52 children with ASD is 0.941 with a 95\% confidence interval of 0.912 to 0.961 (p$<$0.001).
One child in S2 has been excluded from the following analysis due to the limited number of utterances.


\begin{table}[t]
    \centering
    \small
    \begin{tabular}{|c|c|c|c|c|}
        \hline
         \textbf{Group} &  \textit{\textbf{N}}  &  \begin{tabular}[c]{@{}c@{}}\textbf{mean CA} \\ (year;month) \end{tabular}  & \begin{tabular}[c]{@{}c@{}}\textbf{Dur.-all} \\ \textbf{(\textit{N} of utt.)}\end{tabular}   &  \begin{tabular}[c]{@{}c@{}} \textbf{Dur.-ch.} \\ \textbf{(\textit{N} of utt.)} \end{tabular} \\ \hline
         \textbf{ASD}   &  73                   &  7;8              &  \begin{tabular}[c]{@{}c@{}}58h 22m \\ (57,865) \end{tabular} & \begin{tabular}[c]{@{}c@{}}14h 17m \\ (20,841) \end{tabular} \\ \hline
         \textbf{TD}    &  9                    &  7;10             &  \begin{tabular}[c]{@{}c@{}}6h 42m \\ (7,032) \end{tabular} & \begin{tabular}[c]{@{}c@{}}1h 16m \\ (2,231) \end{tabular} \\ \hline
         \multicolumn{5}{r}{\scriptsize \textit{N for the number; CA for chronological age}} \\ 
         \multicolumn{5}{r}{\scriptsize \textit{Dur-all for the entire corpus; Dur-ch. for children's recordings}} \\ 
    \end{tabular}
    \caption{Basic information of the current version of the corpus}
    \label{corpus_table}
\end{table}


\begin{table}[t]
    \centering
    \begin{tabular}{| >{\centering\arraybackslash}m{1cm} | >{\centering\arraybackslash}m{0.5cm} | >{\centering\arraybackslash}m{0.5cm} | >{\centering\arraybackslash}m{0.5cm} | >{\centering\arraybackslash}m{0.5cm} | >{\centering\arraybackslash}m{0.6cm}| >{\centering\arraybackslash}m{0.8cm} | }
        \cline{1-7}
        \multirow{2}{*}{} & \multirow{2}{*}{\textbf{TD}}    &\multicolumn{5}{c|}{\textbf{ASD}}                \\ \cline{3-7}
                          &                                 & \textbf{1} & \textbf{2} &\textbf{3} &\textbf{excl.}
                          & \textbf{Total} \\ \cline{1-7}
        \textbf{SCS}      & \multirow{2}{*}{9}              &  25        &   25       & 23        &   -     & \textbf{73}              \\ \cline{1-1} \cline{3-7}
        \textbf{PP}       &                                 &  15        &   23       & 14        &  21     & \textbf{73}       \\ \hline
        \multicolumn{7}{r}{\small \textit{excl. denotes exclusion due to limited amount}} \\
        \multicolumn{7}{r}{\small \textit{of speech production}}
    \end{tabular}
    \caption{The number of children in each group divided by social communication severity (SCS) levels and pronunciation proficiency (PP) scores}
    \label{group_info_table}
\end{table}

\subsection{Acoustic Analysis}
\textbf{\textit{Acoustic features}}
Various low-level acoustic features are extracted for analyzing the speech of children with ASD in terms of SCS and PP.
The PP score would be directly related to the pronunciation features such as the percentage of correctly pronounced phonemes or vowel space-related features.
However, features pertaining to segmental errors cannot be extracted currently since the speech is not orthographically transcribed. 
As a preliminary study for automatic assessment models, low-level acoustic features are selected.

The feature set employed in the previous study \cite{lee_knowledge-driven_2023}, along with features related to cepstral prominence peak (CPP), are extracted.
The feature set proved effective in capturing the speech characteristics of children with ASD, distinguishing them from TD children. 
The feature set encompasses 5 categories of acoustic features: pitch, spectrum, speech rate, voice quality, and CPP. 
The features within each category are as follows:

\begin{itemize}
\item Pitch: the mean, standard deviation (SD), maximum, minimum, median, 25th percentile, and 75th percentile values of F0s
\item Spectrum: the 12-dimensional MFCCs and a log energy
\item Speech rate: total duration, pause duration, speaking duration, speaking rate, articulation rate, average syllable duration, the number of pauses, and the ratio of speech duration and total duration
\item Voice quality: jitter, shimmer, HNR, the number of voice breaks (VBs), and the percentage of VBs
\item CPP: CPP with voice detection (VD), CPP without VD
\end{itemize}

Mel-frequency cepstral coefficients (MFCCs) provide an alternative feature set for capturing features related to pronunciation. 
The low-order coefficients represent the vocal tract configuration, which is associated with articulation. 
Therefore, log energy and 12-dimensional MFCCs are extracted for the analysis.
CPP features are a more reliable measure of voice quality and can be applied to running speech samples \cite{heman-ackah_cepstral_2003}. 
Spectrum features are extracted using the Librosa library \cite{mcfee_librosalibrosa_2022} and pitch, voice quality, and speech rate features are extracted using the Parselmouth library \cite {jadoul_introducing_2018} in Python. CPP features are extracted through Praat \cite{boersma_praat_nodate}.

\begin{table*}[hbt!]
    \centering
    \scriptsize
    \setlength{\tabcolsep}{5pt}
    \begin{tabularx}{\textwidth}{|c|c|c|c|c|c|c|c|c|c|c|c|}
    \hline
    \multicolumn{2}{|c|}{}      & \multicolumn{5}{c|}{\textbf{Social Communication Severity}}       & \multicolumn{5}{c|}{\textbf{Pronunciation Proficiency}} \\ \cline{3-12}
    \multicolumn{2}{|c|}{\textbf{Features}}  & \multicolumn{3}{c|}{\textbf{inter-group}} & \multicolumn{2}{c|}{\textbf{correlation}}  & \multicolumn{3}{c|}{\textbf{inter-group}} & \multicolumn{2}{c|}{\textbf{correlation}} \\ \cline{3-12}
    \multicolumn{2}{|c|}{}                  & \textbf{Statistics}  & \textbf{\textit{p}}  &  \textbf{post-hoc} & \textbf{$\rho$}  & \textbf{\textit{p}}    & \textbf{Statistics}  & \textbf{\textit{p}}  & \textbf{post-hoc}   & \textbf{$\rho$}  & \textbf{\textit{p}}    \\ \cline{1-12}
    \multirow{13}{*}{Spectrum} & log energy 
            & \textit{F}=0.197  & 0.898 & - & $-$0.035  & 0.757    
            & \textit{F}=0.676  & 0.570 & - & 0.209  & 0.137 \\ \cline{2-12}
        & 1st MFCC 
            & \textit{F}=\textbf{2.920}  & \textbf{0.039\textsuperscript{*}} &-& $-$0.252  & \textbf{0.023\textsuperscript{*}}   
            & \textit{H}=7.374  & 0.061 & - & 0.129  & 0.363  \\ \cline{2-12}
        & 2nd MFCC 
            & \textit{F}=0.169  & 0.917 & - & 0.048  & 0.670    
            & \textit{F}=\textbf{4.234}  & \textbf{0.009\textsuperscript{**}}  & a$>$b & 0.039  & 0.782 \\ \cline{2-12}
        & 3rd MFCC 
            & \textit{F}=0.608  & 0.612  &-& 0.094  & 0.401   
            & \textit{H}=\textbf{17.394}  & \textbf{0.001\textsuperscript{**}} & a$<$b, c & \textbf{0.289}  & \textbf{0.038\textsuperscript{*}} \\ \cline{2-12}
        & 4th MFCC 
            & \textit{F}=0.771  & 0.514 & - & $-$0.091  & 0.418    
            & \textit{F}=2.140  & 0.105 & - & 0.062  & 0.664 \\ \cline{2-12}
        & 5th MFCC 
            & \textit{F}=\textbf{5.109}  & \textbf{0.003\textsuperscript{**}}  &b<c,d& \textbf{0.344}  & \textbf{0.001\textsuperscript{**}}    
            & \textit{H}=5.978  & 0.113 & - & 0.233  & 0.097 \\ \cline{2-12}
        & 6th MFCC 
            & \textit{F}=1.193 & 0.318  & - & 0.046  & 0.684   
            & \textit{H}=7.354  & 0.061 & - & $-$0.032  & 0.822  \\ \cline{2-12}
        & 7th MFCC 
            & \textit{F}=\textbf{6.241}  & \textbf{0.001\textsuperscript{**}}    &b<c,d&     0.310  & 0.005   
            & \textit{H}=5.669  & 0.129 & - & \textbf{0.296}  & \textbf{0.033\textsuperscript{*}} \\ \cline{2-12}
        & 8th MFCC 
            & \textit{F}=0.952  & 0.420 & - & 0.145  & 0.195   
            & \textit{F}=\textbf{9.241}  & \textbf{0.000\textsuperscript{***}} & a$<$b,c,d & $-$0.039  & 0.784 \\ \cline{2-12}
        & 9th MFCC 
            & \textit{F}=0.960 & 0.416  &-& 0.156  & 0.162    
            & \textit{H}=5.840  & 0.120 &-& 0.203  & 0.148 \\ \cline{2-12}
        & 10th MFCC 
            & \textit{F}=1.056  & 0.372 &-& 0.258  & \textbf{0.019\textsuperscript{*}}    
            & \textit{F}=\textbf{3.547}  & \textbf{0.020\textsuperscript{*}} & a$<$b,c & $-$0.197  & 0.162 \\ \cline{2-12}
        & 11th MFCC 
            & \textit{H}=1.677  & 0.642 &-& 0.103  & 0.355   
            & \textit{H}=6.336  & 0.096 &-& $-$0.006  & 0.966 \\ \cline{2-12}
        & 12th MFCC 
            & \textit{F}=\textbf{4.209}  & \textbf{0.008\textsuperscript{**}} &b<c& 0.248  & \textbf{0.025\textsuperscript{*}}   
            & \textit{H}=\textbf{7.870}  & \textbf{0.049\textsuperscript{*}}  &-& \textbf{0.313}  & \textbf{0.024\textsuperscript{*}}  \\ \hline
    \multirow{16}{*}{\begin{tabular}[c]{@{}c@{}}Speech\\ Rate\end{tabular}} & \begin{tabular}[c]{@{}c@{}}total\\ duration\end{tabular} 
            & \textit{H}=5.830  & 0.120 &-& 0.023  & 0.835   
            & \textit{H}=1.564  & 0.668 &-& $-$0.082  & 0.563 \\ \cline{2-12}
        & \begin{tabular}[c]{@{}c@{}}speech\\ duration\end{tabular} 
            & \textit{F}=1.540  & 0.211 &-& 0.099  & 0.374   
            & \textit{F}=2.040  & 0.118 &-& $-$0.084  & 0.553 \\ \cline{2-12}
        & \begin{tabular}[c]{@{}c@{}}speaking\\ rate\end{tabular} 
            & \textit{F}=1.709  & 0.172 &-& 0.059  & 0.597   
            & \textit{F}=\textbf{5.762}  & \textbf{0.002\textsuperscript{**}} & b$<$c & \textbf{0.328}  & \textbf{0.018\textsuperscript{*}} \\ \cline{2-12}
        & \begin{tabular}[c]{@{}c@{}}articulation\\ rate\end{tabular} 
            & \textit{H}=3.706  & 0.295  &-& 0.138  & 0.217  
            & \textit{H}=\textbf{14.203}  & \textbf{0.003\textsuperscript{**}} & b$<$c & 0.235  & 0.094 \\ \cline{2-12}
        & \begin{tabular}[c]{@{}c@{}}\textit{N}. of\\ pauses\end{tabular} 
            & \textit{H}=0.299  & 0.393 &-& 0.138  & 0.216   
            & \textit{F}=0.401  & 0.753 &-& $-$0.126  & 0.374 \\ \cline{2-12}
        & \begin{tabular}[c]{@{}c@{}}avrg. syl.\\ dur.\end{tabular} 
            & \textit{F}=0.869  & 0.461 &-& $-$0.143  & 0.200   
            & \textit{H}=\textbf{10.142}  & \textbf{0.017\textsuperscript{*}} & b$>$c & $-$0.202  & 0.151 \\ \cline{2-12}
        & \begin{tabular}[c]{@{}c@{}}phonation\\ ratio\end{tabular} 
            & \textit{H}=5.940  & 0.114 &-& -0.075  & 0.504   
            & \textit{H}=7.261  & 0.064 &-& \textbf{0.308}  & \textbf{0.026\textsuperscript{*}} \\ \cline{2-12}
        & \begin{tabular}[c]{@{}c@{}}pause\\ duration\end{tabular} 
            & \textit{H}=4.329  & 0.228 &-& 0.146  & 0.191    
            & \textit{H}=3.003  & 0.391 &-& $-$0.120  & 0.397 \\ \hline
    \multirow{5}{*}{\begin{tabular}[c]{@{}c@{}}Voice\\ Quality\end{tabular}} & jitter 
            & \textit{H}=\textbf{17.763}  & \textbf{0.000\textsuperscript{***}} &b<c,d& \textbf{0.334}  & \textbf{0.002\textsuperscript{**}}  
            & \textit{H}=\textbf{12.914}  & \textbf{0.005\textsuperscript{**}} & b$<$d  & \textbf{0.493}  & \textbf{0.000\textsuperscript{***}}  \\ \cline{2-12}
        & shimmer 
            & \textit{H}=3.557  & 0.313   &-& 0.124  & 0.265  
            & \textit{H}=\textbf{8.147}  & \textbf{0.043\textsuperscript{*}}  & b$<$c & \textbf{0.275}  & \textbf{0.048\textsuperscript{*}}\\ \cline{2-12}
        & HNR 
            & \textit{H}=2.955  & 0.399  &-& -0.077  & 0.490   
            & \textit{H}=\textbf{8.777}  & \textbf{0.032\textsuperscript{*}} & b$>$c & \textbf{$-$0.331}  & \textbf{0.016\textsuperscript{*}} \\ \cline{2-12}
        & \textit{N}. of VBs 
            & \textit{F}=\textbf{3.206}  & \textbf{0.028\textsuperscript{*}} &-& \textbf{0.259}  & \textbf{0.019\textsuperscript{*}}  
            & \textit{F}=1.905  & 0.139  &-& 0.100  & 0.478 \\ \cline{2-12}
        & \%VBs 
            & \textit{H}=\textbf{16.916}  & \textbf{0.001\textsuperscript{**}} &b<d& \textbf{0.357} & \textbf{0.001\textsuperscript{***}}   
            & \textit{H}=5.392  & 0.145 &-& 0.259  & 0.064 \\ \hline
    \multirow{5}{*}{Pitch} & median 
            & \textit{F}=\textbf{5.289}  & \textbf{0.002\textsuperscript{**}} &b,c<d& \textbf{0.315}  & \textbf{0.004\textsuperscript{**}}  
            & \textit{F}=\textbf{3.031}  & \textbf{0.037\textsuperscript{*}}  & - & 0.020  & 0.890  \\ \cline{2-12}
        & mean 
            & \textit{F}=\textbf{5.236}  & \textbf{0.002\textsuperscript{**}} &b,c<d& \textbf{0.320}  & \textbf{0.003\textsuperscript{**}}    
            & \textit{F}=2.758  & 0.050  &-& 0.044  & 0.756  \\ \cline{2-12}
        & SD 
            & \textit{F}=\textbf{5.870}  & \textbf{0.001\textsuperscript{**}} &b<c,d& \textbf{0.323}  & \textbf{0.003\textsuperscript{**}}    & \textit{F}=\textbf{4.092}  & \textbf{0.011\textsuperscript{*}} & a$<$d & 0.254  & 0.069  \\ \cline{2-12}
        & minimum 
            & \textit{H}=1.267  & 0.737 &-& 0.099  & 0.377   
            & \textit{H}=1.593  & 0.661 &-& $-$0.025  & 0.862 \\ \cline{2-12}
        & maximum 
            & \textit{F}=\textbf{5.937}  & \textbf{0.001\textsuperscript{**}} &b<d& \textbf{0.374}  & \textbf{0.000\textsuperscript{***}}   
            & \textit{F}=4.030  & \textbf{0.011\textsuperscript{*}} &-& 0.200  & 0.155  \\ \hline
    \multirow{3}{*}{CPP} & \begin{tabular}[c]{@{}c@{}}CPP w/\\ VD\end{tabular} 
            & \textit{F}=\textbf{8.484}  & \textbf{0.000\textsuperscript{***}} &b>c,d& \textbf{$-$0.417}  & \textbf{0.000\textsuperscript{***}}   
            & \textit{H}=\textbf{12.810}  & \textbf{0.005\textsuperscript{**}} & b,c$>$d & \textbf{$-$0.485}  & \textbf{0.000\textsuperscript{***}} \\ \cline{2-12}
        & \begin{tabular}[c]{@{}c@{}}CPP w/o\\ VD\end{tabular} 
            & \textit{H}=\textbf{25.120}  & \textbf{0.000\textsuperscript{***}} &b,c>d& \textbf{$-$0.446}  & \textbf{0.000\textsuperscript{***}}    
            & \textit{H}=\textbf{11.555}  & \textbf{0.009\textsuperscript{**}} & b,c$>$d & \textbf{$-$0.399}  & \textbf{0.003\textsuperscript{**}} \\ \hline
    \multicolumn{12}{r}{Abbreviation: VBs for voice breaks; SD for standard deviation; VD for voice detection} \\
    \multicolumn{12}{r}{\textit{F} in statistic resulted from one-way ANOVA test and \textit{H} from Kruskal Wallis test} \\
    \multicolumn{12}{r}{a: TD, b: score 1, c: score 2, d: score 3} \\
    \multicolumn{12}{r}{*\textit{p}$<$0.05, **\textit{p}$<$0.01, ***\textit{p}$<$0.001} \\
    \end{tabularx}
    \caption{Analysis results of acoustic features for inter-group differences among TD and three ASD subgroups and correlation with clinical scores}
    \label{acoustic_result_table}
\end{table*}

\textbf{\textit{Analysis methods}}
The acoustic features are extracted from the speech data from the current version of the corpus. 
The feature values are averaged for each child because the SCS levels and PP scores are rated at the speaker level. 

A comparative analysis is carried out among the 4 groups: the three ASD groups and the TD group.
In cases where both the assumptions of normality and homogeneous variance are met, a one-way ANOVA is performed, followed by Bonferroni post hoc tests. 
Alternatively, if either normality assumption or homogeneity of variance assumption is violated, a Kruskal-Wallis test is conducted, followed by Dunn's post hoc test. 
These statistical analyses are conducted using SciPy library \cite{virtanen_scipy_2020} and Scikit-posthocs \cite{terpilowski_scikit-posthocs_2019} in Python.
Then the same feature values are normalized to have a zero mean and unit variance to be examined if each of them has a significant correlation with SCS levels and PP scores.
Spearman's rank correlation coefficients are computed in Python using SciPy library \cite{virtanen_scipy_2020}, as the feature values are continuous whereas scores are categorical.
The Spearman's rank correlation coefficient $\rho$ between pronunciation and each feature is calculated using SciPy library \cite{virtanen_scipy_2020} in Python.

\textbf{\textit{Results}}
The results of multiple group comparisons and post hoc tests are detailed in Table \ref{acoustic_result_table}. 
This table also presents Spearman's rank correlation coefficients between each feature and the SCS levels or PP scores, accompanied by their respective p-values.
The comparative analysis shows that certain acoustic features display differences within ASD subgroups categorized by SCS level, though no acoustic feature distinguishes TD from any ASD subgroups.
In case of MFCCs, the 5th, 7th, and 12th MFCCs of L1 are different from L2, and all of them except for the 12th MFCC also differ from L3.
Regarding voice quality features, jitter and the percentage of voice breaks exhibit significant differences within ASD subgroups. 
Pitch-related and CPP features show significant variations within these subgroups. 

In contrast, distinct patterns emerge for PP scores. Some acoustic features show differences between TD children and ASD subgroups categorized by PP score.
The 2nd, 3rd, 8th, and 10th MFCCs of TD are different from those of S1. All of them except for the 2nd MFCC also differ from S2. The 8th MFCC is different from that in S3. 
Among the 3 subgroups of ASD, features in voice quality and CPP differed.
It implies that voice quality-related features would play a major role in the perceptual pronunciation proficiency evaluation of children with ASD.

Correlation analyses between acoustic features and SCS levels indicate that voice quality-related features, especially CPPs, have a moderate relationship with SCS levels. This suggests that children with ASD exhibiting higher SCS levels might demonstrate poorer vocal control. 
The correlation analysis based on PP scores presents that voice quality-related features, such as jitter and CPPs, are moderately related to the scores. 
The positive coefficient $\rho$ of jitter corresponds to its general interpretation, as higher jitter is associated with greater variance in the voice, leading to poorer speech quality. 
In contrast, the correlation is negative for CPP features. It is because a lower CPP value is linked with poorer vocal control \cite{heman-ackah_cepstral_2003}.
Notably, they are the features which show significant differences within the ASD groups. It further supports the role of voice quality-related features in pronunciation evaluation.


\subsection{Linguistic Analysis}

\textbf{\textit{Linguistic features}}
Since SCS levels are closely related to children's social interactions, we aim to capture their characteristics not only from acoustic features but also from linguistic features. 

To extract linguistic features, The Linguistic Feature Toolkit (LFTK) \cite{lee-lee-2023-lftk} is employed. 
The LFTK is a comprehensive toolkit that comprises over 200 handcrafted features collected and categorized from previous research in areas such as text readability assessment, automated essay scoring, fake news detection, and paraphrase detection. 
We carefully select a linguistic feature set that takes into consideration the data collection process and the specific characteristics of the Korean language.

\begin{table*}[hbt!]
    \centering
    \footnotesize
    \setlength{\tabcolsep}{7.64pt}
    \resizebox{\textwidth}{!}{%
    \begin{tabularx}{\textwidth}{|c|c|c|c|c|c|c|}
    \hline
    \multicolumn{2}{|c|}{\multirow{3}{*}{\textbf{Features}}} & \multicolumn{5}{c|}{\textbf{Social Communication Severity}} \\ \cline{3-7}
    \multicolumn{2}{|c|}{} & \multicolumn{3}{c|}{\textbf{inter-group}}  & \multicolumn{2}{c|}{\textbf{correlation}} \\ \cline{3-7}
    \multicolumn{2}{|c|}{} & \textbf{Statistics} & \textbf{\textit{p}} & \textbf{post-hoc} & \textbf{$\rho$}  & \textbf{\textit{p}}  \\ \cline{1-7}
    \multicolumn{1}{|c|}{\multirow{5}{*}{Wordsent}}                & total \# of words              
        & \textit{H}=\textbf{19.772}    & \textbf{0.000\textsuperscript{***}}   & a$>$d& \textbf{$-$0.410}         & \textbf{0.000\textsuperscript{***}} \\ \cline{2-7} 
    ~ & total \# of stop words
        & \textit{H}=\textbf{14.625}    & \textbf{0.002\textsuperscript{**}}    &a$>$d& \textbf{$-$0.341}         & \textbf{0.002\textsuperscript{**}} \\ \cline{2-7} 
    ~ & total \# of syllables
        & \textit{F}=1.033              & 0.382                                 &-& \textbf{$-$0.326}         & \textbf{0.003\textsuperscript{**}} \\ \cline{2-7} 
    ~ & total \# of unique words       
        & \textit{H}=\textbf{26.131}    & \textbf{0.000\textsuperscript{***}}   &b$>$c,d& \textbf{$-$0.458}         & \textbf{0.000\textsuperscript{***}} \\ \cline{2-7} 
    ~ & total \# of sent.        
        & \textit{H}=\textbf{30.757}    & \textbf{0.000\textsuperscript{***}}   &b$>$c,d& \textbf{$-$0.501}         & \textbf{0.000\textsuperscript{***}} \\ \hline
    \multicolumn{1}{|c|}{\multirow{10}{*}{Part of Speech}}         & total \# of adj.               
        & \textit{H}=\textbf{22.939}    & \textbf{0.000\textsuperscript{***}}   &b$>$c,d& \textbf{$-$0.427}         & \textbf{0.000\textsuperscript{***}} \\ \cline{2-7} 
    ~ & total \# of adp.               
        & \textit{F}=2.068              & 0.111                                 &-& $-$0.167                  & 0.131 \\ \cline{2-7} 
    ~ & total \# of adv.               
        & \textit{F}=\textbf{3.901}     & \textbf{0.012\textsuperscript{**}}    &b$>$d& \textbf{$-$0.309}         & \textbf{0.004\textsuperscript{**}} \\ \cline{2-7} 
    ~ & total \# of nouns              
        & \textit{F}=\textbf{4.392}     & \textbf{0.006\textsuperscript{**}}    &b$>$d& \textbf{$-$0.413}         & \textbf{0.000\textsuperscript{***}} \\ \cline{2-7} 
    ~ & total \# of verbs              
        & \textit{H}=\textbf{29.492}    & \textbf{0.000\textsuperscript{***}}   &b$>$c,d& \textbf{$-$0.473}         & \textbf{0.000\textsuperscript{***}} \\ \cline{2-7} 
    ~ & total \# of unique adj.        
        & \textit{H}=\textbf{26.697}    & \textbf{0.000\textsuperscript{***}}  &b$>$c,d& \textbf{$-$0.466}         & \textbf{0.000\textsuperscript{***}} \\ \cline{2-7} 
    ~ & total \# of unique adp.        
        & \textit{H}=\textbf{12.771}    & \textbf{0.005\textsuperscript{**}}    &b$>$d& \textbf{$-$0.314}         & \textbf{0.004\textsuperscript{**}} \\ \cline{2-7} 
    ~ & total \# of unique adv.        
        & \textit{H}=\textbf{22.681}    & \textbf{0.000\textsuperscript{***}}  &b$>$c,d& \textbf{$-$0.423}         & \textbf{0.000\textsuperscript{***}} \\ \cline{2-7} 
    ~ & total \# of unique nouns       
        & \textit{H}=\textbf{21.996}    & \textbf{0.000\textsuperscript{***}}  & b$>$d& \textbf{$-$0.429}         & \textbf{0.000\textsuperscript{***}}\\ \cline{2-7} 
    ~ & total \# of unique verbs       
        & \textit{H}=\textbf{31.176}   & \textbf{0.000\textsuperscript{***}}   &b$>$c,d& \textbf{$-$0.491}         & \textbf{0.000\textsuperscript{***}} \\ \hline
    \multicolumn{1}{|c|}{Average Wordsent}                         & avrg. \# of words per sent. 
        & \textit{F}=2.480              & 0.067                                &-& \textbf{0.356}          & \textbf{0.001\textsuperscript{***}}  \\ \hline
    \multicolumn{1}{|c|}{\multirow{10}{*}{\begin{tabular}[c]{@{}c@{}}Average\\ Part of Speech\end{tabular}}} & avrg. \# of adj. per word      
        & \textit{F}=1.185              & 0.321                                 &-& $-$0.187                          & 0.090                                      \\ \cline{2-7} 
    ~ & avrg. \# of adp. per word      
        & \textit{F}=\textbf{2.779}  & \textbf{0.046\textsuperscript{*}} &b$<$d& \textbf{0.235}  & \textbf{0.032\textsuperscript{*}} \\ \cline{2-7} 
    ~ & avrg. \# of adv. per word      
        & \textit{H}=2.015  & 0.569    &-& 0.014  & 0.901 \\ \cline{2-7} 
    ~ & avrg. \# of nouns per word     
        & \textit{F}=1.288  & 0.284 &-& $-$0.070  & 0.530\\ \cline{2-7} 
    ~ & avrg. \# of verbs per word 
        & \textit{H}=\textbf{25.859} & \textbf{0.000\textsuperscript{***}} &b$>$c,d& \textbf{$-$0.468}  & \textbf{0.000\textsuperscript{***}} \\ \cline{2-7} 
    ~ & avrg. \# of adj. per sent.  
        & \textit{F}=1.877  & 0.140 &-& 0.133  & 0.229  \\ \cline{2-7} 
    ~ & avrg. \# of adp. per sent.  
        & \textit{F}=1.894  & 0.137 &-& \textbf{0.329} & \textbf{0.002\textsuperscript{**}}  \\ \cline{2-7} 
    ~ & avrg. \# of adv. per sent. 
        & \textit{H}=6.998  & 0.072 &-& \textbf{0.247}  & \textbf{0.024\textsuperscript{*}}   \\ \cline{2-7} 
    ~ & avrg. \# of nouns per sent. 
        & \textit{F}=2.118  & 0.104 &-& \textbf{0.246}  & \textbf{0.025\textsuperscript{*}} \\ \cline{2-7} 
    ~ & avrg. \# of verbs per sent. 
        & \textit{F}=1.189  & 0.319 &-& $-$0.039  & 0.725  \\ \hline
    \multicolumn{1}{|c|}{\multirow{5}{*}{Lexical Variation}}       & simple adj. variation          
        & \textit{F}=0.511               & 0.676                                &-& 0.048                              & 0.663  \\ \cline{2-7} 
    ~ & simple adp. variation          
        & \textit{F}=1.628  & 0.189 &-& \textbf{$-$0.229}  & \textbf{0.037\textsuperscript{*}} \\ \cline{2-7} 
    ~ & simple adv. variation          
        & \textit{H}=0.972  & 0.808 &-& $-$0.039  & 0.723  \\ \cline{2-7} 
    ~ & simple noun variation          
        & \textit{F}=0.550  & 0.650 &-& 0.084  & 0.450  \\ \cline{2-7} 
    ~ & simple verb variation          
        & \textit{F}=0.015  & 0.997 &-& 0.111  & 0.316 \\ \hline
    \multicolumn{1}{|c|}{Type Token Ratio} & simple TTR                     
        & \textit{F}=0.104  & 0.957 &-& $-$0.082  & 0.459 \\ \hline
    \multicolumn{7}{r}{Abbreviation: sent. for sentence; adj. for adjective; adp. for adposition; adv for adverb} \\ 
    \multicolumn{7}{r}{\textit{F} in statistics resulted from one-way ANOVA test and \textit{H} from Kruskal Wallis test} \\     
    \multicolumn{7}{r}{a: TD, b: level 1, c: level 2, d: level 3} \\
    \multicolumn{7}{r}{*\textit{p}$<$0.05, **\textit{p}$<$0.01, ***\textit{p}$<$0.001} \\

    \end{tabularx} %
    }
    \caption{Analysis results of linguistic features for inter-group differences among TD and three ASD subgroups and correlation with social communication severity levels}
    \label{text_result_table}
\end{table*}

The linguistic feature set encompasses six categories of features, each designed to capture various aspects of the text data:
\begin{itemize}
    \item Wordsent: This category includes basic word and sentence counts, such as the total number of words, total number of stop words, total number of syllables, total number of unique words, and total number of sentences.
    \item Partofspeech: These features pertain to part-of-speech properties and cover aspects like the total number of adjectives, adpositions, adverbs, nouns, and verbs. It also includes counts of unique adjectives, adpositions, adverbs, nouns, and vowels.
    \item Avgwordsent: This category calculates the average number of words per sentence.
    \item Avgpartofspeech: It calculates averages related to part-of-speech features, including the average number of adjectives per word, adpositions per word, adverbs per word, nouns per word, verbs per word, adjectives per sentence, adpositions per sentence, adverbs per sentence, nouns per sentence, and verbs per sentence.
    \item Lexicalvariation: These features measure lexical variation and include variations in simple adjectives, adpositions, adverbs, nouns, and verbs.
    \item Typetokenratio: The type token ratio is used to capture the lexical richness of a text.
\end{itemize}
Each linguistic feature value is computed based on the transcription, which transcribes the entire recording session.

\textbf{\textit{Analysis methods}}
Linguistic analysis is conducted using the same methodology as acoustic analysis.

\textbf{\textit{Results}}
The results of both comparative and correlation analyses are provided in Table \ref{text_result_table}.
Significant differences emerged across the four groups. 
Within the Wordsent features, key parameters like the total number of words, stop words, unique words, and sentences showed significant variations across groups. Significant differences were observed between L1 and L3 in terms of these parameters, emphasizing linguistic differences in children with varying social communication severity. 
Partofspeech features revealed significant variations, with post hoc analyses identifying specific group pairs with notable differences. 

In the correlation analysis involving linguistic features and SCS levels, features reflecting a child's active participation, such as the total number of sentences and words, show moderate negative correlations with SCS levels. This suggests that children with higher SCS levels tend to participate less in sessions, resulting in limited verbal communication. Notably, more linguistic features show significant correlations with SCS levels compared to acoustic features, emphasizing the importance of linguistic analysis in examining children with ASD's speech. The relationships between linguistic features and social communication are complex, but these results offer insights into potential linguistic markers for further exploration in clinical contexts.

\subsection{Discussion}

The outcomes from statistical analyses point out distinct speech characteristics in children with ASD compared to TD children. 
It is notable that MFCCs are the major features differentiating TD and ASD subgroups divided by PP scores. 
Considering that MFCCs are more related to pronunciation than other features, it could reflect the poorer articulatory skills in children with ASD as reported in previous studies on their speech production \cite{rapin_subtypes_2009, cleland_phonetic_2010, shriberg_hypothesis_2011, wolk_phonological_2013}.
This finding also endorse the necessity for an analysis of pronunciation-related features in speech.

The features exhibiting differences between children with ASD and TD in this study do not overlap with the study on the differentiation of children with ASD from TD \cite{lee_knowledge-driven_2023}.
It can be attributed to methodological differences, as this study computes group averages on a per-speaker basis, while \citet{lee_knowledge-driven_2023} computed group averages based on utterances.

Voice quality-related features show not only differences across subgroups of ASD but also correlation with SCS levels and PP scores. 
It is consistent with \citet{bone_spontaneous-speech_2012} which revealed the relation between voice quality features and the degree of atypicality of speech in children with ASD.
It is worth noting the relationship between SCS and voice quality.
It could be explained by their difficulties in controlling vocal tones as reported in \citet{wetherby2006understanding}.
Voice quality-related features could serve as indicators of the challenges these children face in effective communication and social engagement.

There were differences between results regarding SCS levels and those regarding PP scores. 
While MFCCs differed among the subgroups of ASD categorized by SCS levels, they differentiated ASD subgroups from TD from the perspective of pronunciation. 
Regarding the differences among the subgroups, pitch features exhibited variations in relation to SCS levels, whereas speech rate features differed among the subgroups based on PP scores.
In addition to the acoustic features, linguistic features also demonstrated correlations with SCS levels, indicating the degree of participation.
These contrasting results highlight the fact that assessing SCS and PP is based on distinct aspects of speech, emphasizing the multifaceted nature of speech evaluation in children with ASD.

Some features did not differ among the groups or showed a minor or negligible correlation with the clinical scores. 
Due to the substantial variability in speech characteristics among children with ASD, it is plausible that features could overlap within the subgroups.
However, these features could still hold usefulness for developing an automatic evaluation model in that features utilized to train a classification model did not necessarily coincide with those showing significant differences between children with ASD and TD \cite{lee_knowledge-driven_2023}.
The influence of each feature should be identified in further research.

\section{Conclusion}
A speech corpus of Korean children with ASD, which provides clinical scores on SCS and PP, is constructed for the first time and analyzed for automatic assessment systems for social communication and pronunciation. 
To reflect speech and linguistic aspects in children with ASD, interactions during speech and language evaluation sessions between children with ASD and SLPs were recorded. 
The speech corpus is to be composed of speech recordings collected from 300 children with ASD and 50 children with TD, with the current version containing speech from 73 children with ASD and 9 TD children.

To explore the acoustic and linguistic characteristics in relation to the clinical scores, comparative analyses, and correlation analyses are performed.
The children with ASD are divided into three subgroups based on their SCS levels and PP scores, respectively, and the differences in the acoustic features among the three ASD groups and TD children are examined. 
As a result, features within MFCCs, voice quality, and pitch are different among the groups divided by the SCS levels. Certain features within MFCCs, speech rate, voice quality, and pitch exhibit significant differences across the groups divided by PP scores.
Voice quality-related features show a significant moderate correlation with SCS levels and PP scores.
SCS levels are also in significant moderate correlations with linguistic features embracing various surface lexical features. 
These findings underscore the multifaceted nature of evaluation for children with ASD and are also invaluable for clinicians and researchers working to enhance our understanding of ASD and develop effective automatic assessment systems.

One limitation of these analyses is the absence of features directly associated with SCS and PP.
At this moment, it is unable to identify the exact content of the utterances and mispronounced speech sounds as it is phonemically transcribed.
Once orthographic transcriptions are provided, the analyses will be extended with the PP- and SCS-related features.
Future work involves developing automatic evaluation models for SCS and PP, reflecting the analysis findings.

\section{Acknowledgements}
Funding and IRB approval are hidden at the moment.

\section{Bibliographical References}\label{sec:reference}

\bibliographystyle{lrec-coling2024-natbib}
\bibliography{o-cocosda}

\begin{thebibliography}{42}
\expandafter\ifx\csname natexlab\endcsname\relax\def\natexlab#1{#1}\fi

\bibitem[{Albo-Canals et~al.(2013)Albo-Canals, Heerink, Diaz, Padillo, Maristany, Barco, Angulo, Riccio, Brodsky, Dufresne et~al.}]{albo2013comparing}
Jordi Albo-Canals, Marcel Heerink, Marta Diaz, Vanesa Padillo, Marta Maristany, Alex Barco, Cecilio Angulo, Ariana Riccio, Lauren Brodsky, Simone Dufresne, et~al. 2013.
\newblock Comparing two lego robotics-based interventions for social skills training with children with asd.
\newblock In \emph{2013 IEEE RO-MAN}, pages 638--643. IEEE.

\bibitem[{{American Psychiatric Association}(2013)}]{american_psychiatric_association_diagnostic_2013}
{American Psychiatric Association}. 2013.
\newblock \emph{Diagnostic and statistical manual of mental disorders}, 5th edition edition.
\newblock American Psychiatric Association.

\bibitem[{Black et~al.(2011)Black, Bone, Williams, Gorrindo, Levitt, and Narayanan}]{black_usc_2011}
Matthew~P. Black, Daniel Bone, Marian~E. Williams, Phillip Gorrindo, Pat Levitt, and Shrikanth Narayanan. 2011.
\newblock The {USC} {CARE} corpus: child-psychologist interactions of children with autism spectrum disorders.
\newblock In \emph{Interspeech 2011}, pages 1497--1500. ISCA.

\bibitem[{Boersma and Weenink(2023)}]{boersma_praat_nodate}
Paul Boersma and David Weenink. 2023.
\newblock \href {http://www.praat.org} {Praat: doing phonetics by computer (version 6.3.10)}.

\bibitem[{Bone et~al.(2012)Bone, Black, Lee, Williams, Levitt, Lee, and Narayanan}]{bone_spontaneous-speech_2012}
Daniel Bone, Matthew~P. Black, Chi-Chun Lee, Marian~E. Williams, Pat Levitt, Sungbok Lee, and Shrikanth Narayanan. 2012.
\newblock Spontaneous-speech acoustic-prosodic features of children with autism and the interacting psychologist.
\newblock In \emph{Interspeech 2012}, pages 1043--1046. {ISCA}.

\bibitem[{Bonneh et~al.(2011)Bonneh, Levanon, Dean-Pardo, Lossos, and Adini}]{bonneh_abnormal_2011}
Yoram~S. Bonneh, Yoram Levanon, Omrit Dean-Pardo, Lan Lossos, and Yael Adini. 2011.
\newblock Abnormal speech spectrum and increased pitch variability in young autistic children.
\newblock \emph{Frontiers in Human Neuroscience}, 4:237.

\bibitem[{Charman et~al.(2005)Charman, Taylor, Drew, Cockerill, Brown, and Baird}]{charman_outcome_2005}
Tony Charman, Emma Taylor, Auriol Drew, Helen Cockerill, Jo-Anne Brown, and Gillian Baird. 2005.
\newblock Outcome at 7 years of children diagnosed with autism at age 2: predictive validity of assessments conducted at 2 and 3 years of age and pattern of symptom change over time.
\newblock \emph{Journal of Child Psychology and Psychiatry}, 46(5):500--513.

\bibitem[{Choi et~al.(2012)Choi, Kim, Kim, Lee, Um, and Chung}]{qolt}
Dae-Lim Choi, Bong-Wan Kim, Yeon-Whoa Kim, Yong-Ju Lee, Yongnam Um, and Minhwa Chung. 2012.
\newblock Dysarthric speech database for development of {Q}o{LT} software technology.
\newblock In \emph{Proceedings of the Eighth International Conference on Language Resources and Evaluation ({LREC}'12)}, pages 3378--3381, Istanbul, Turkey. European Language Resources Association (ELRA).

\bibitem[{Cilia et~al.(2021)Cilia, Carette, Elbattah, Dequen, Gu{\'e}rin, Bosche, Vandromme, Le~Driant et~al.}]{cilia2021computer}
Federica Cilia, Romuald Carette, Mahmoud Elbattah, Gilles Dequen, Jean-Luc Gu{\'e}rin, J{\'e}r{\^o}me Bosche, Luc Vandromme, Barbara Le~Driant, et~al. 2021.
\newblock Computer-aided screening of autism spectrum disorder: Eye-tracking study using data visualization and deep learning.
\newblock \emph{JMIR human factors}, 8(4):e27706.

\bibitem[{Cleland et~al.(2010)Cleland, Gibbon, Peppé, O'Hare, and Rutherford}]{cleland_phonetic_2010}
Joanne Cleland, Fiona~E. Gibbon, Sue J.~E. Peppé, Anne O'Hare, and Marion Rutherford. 2010.
\newblock Phonetic and phonological errors in children with high functioning autism and asperger syndrome.
\newblock \emph{International Journal of Speech-Language Pathology}, 12(1):69--76.

\bibitem[{Clemente(2008)}]{clemente2008recording}
Ignasi Clemente. 2008.
\newblock Recording audio and video.
\newblock \emph{The Blackwell guide to research methods in bilingualism and multilingualism}, pages 177--191.

\bibitem[{Diehl and Paul(2012)}]{diehl_acoustic_2012}
Joshua~John Diehl and Rhea Paul. 2012.
\newblock Acoustic differences in the imitation of prosodic patterns in children with autism spectrum disorders.
\newblock \emph{Research in autism spectrum disorders}, 6(1):123--134.

\bibitem[{Frigaux et~al.(2019)Frigaux, Evrard, and Lighezzolo-Alnot}]{frigaux2019adiados}
A~Frigaux, R~Evrard, and J~Lighezzolo-Alnot. 2019.
\newblock Adi-r and ados and the differential diagnosis of autism spectrum disorders: Interests, limits and openings.
\newblock \emph{L'encephale}, 45(5):441--448.

\bibitem[{Fusaroli et~al.(2017)Fusaroli, Lambrechts, Bang, Bowler, and Gaigg}]{fusaroli_is_2017}
Riccardo Fusaroli, Anna Lambrechts, Dan Bang, Dermot~M. Bowler, and Sebastian~B. Gaigg. 2017.
\newblock Is voice a marker for autism spectrum disorder? a systematic review and meta-analysis.
\newblock \emph{Autism Research}, 10(3):384--407.

\bibitem[{Gale et~al.(2019)Gale, Chen, Dolata, Van~Santen, and Asgari}]{gale_improving_2019}
Robert Gale, Liu Chen, Jill Dolata, Jan Van~Santen, and Meysam Asgari. 2019.
\newblock Improving {ASR} systems for children with autism and language impairment using domain-focused {DNN} transfer techniques: 20th {Annual} {Conference} of the {International} {Speech} {Communication} {Association}: {Crossroads} of {Speech} and {Language}, {INTERSPEECH} 2019.
\newblock \emph{Proceedings of the Annual Conference of the International Speech Communication Association, INTERSPEECH}, 2019-September:11--15.

\bibitem[{Heman-Ackah et~al.(2003)Heman-Ackah, Heuer, Michael, Ostrowski, Horman, Baroody, Hillenbrand, and Sataloff}]{heman-ackah_cepstral_2003}
Yolanda~D. Heman-Ackah, Reinhardt~J. Heuer, Deirdre~D. Michael, Rosemary Ostrowski, Michelle Horman, Margaret~M. Baroody, James Hillenbrand, and Robert~T. Sataloff. 2003.
\newblock Cepstral peak prominence: a more reliable measure of dysphonia.
\newblock \emph{The Annals of Otology, Rhinology, and Laryngology}, 112(4):324--333.

\bibitem[{Jadoul et~al.(2018)Jadoul, Thompson, and Boer}]{jadoul_introducing_2018}
Yannick Jadoul, Bill Thompson, and Bart~de Boer. 2018.
\newblock Introducing parselmouth: A python interface to praat.
\newblock \emph{Journal of Phonetics}, 71:1--15.

\bibitem[{Kim et~al.(2007)Kim, Pae, and Park}]{kim_assessment_2007}
Min~Jung Kim, Soyeong Pae, and Chang~Il Park. 2007.
\newblock \emph{Assessment of Phonology and Articulation for Children ({APAC})}.
\newblock Human Brain Research \& Counseling.

\bibitem[{Kim et~al.(2009)Kim, Hong, KyungHee, Hae-Seong, and Lee}]{revt}
Young~Tae Kim, Gyung~Hun Hong, Kim KyungHee, Chang Hae-Seong, and Ju-Yeon Lee. 2009.
\newblock \emph{Receptive \& expressive vocabulary test ({REVT})}.
\newblock Seoul Community Rehabilitation Center.

\bibitem[{Kim et~al.(2003{\natexlab{a}})Kim, Kim, Yoon, and Kim}]{selsi}
Young~Tae Kim, KyungHee Kim, Hea~Ryun Yoon, and Wha-soo Kim. 2003{\natexlab{a}}.
\newblock \emph{Sequenced Language Scale for Infants ({SELSI})}.
\newblock Special Education Publishing.

\bibitem[{Kim et~al.(2003{\natexlab{b}})Kim, Seong, and Lee}]{pres}
Young~Tae Kim, Tae-Je Seong, and YoonKyoung Lee. 2003{\natexlab{b}}.
\newblock \emph{Preschool Receptive-Expressive Language Scale ({PRES})}.
\newblock Seoul Community Rehabilitation Center.

\bibitem[{Kojovic et~al.(2021)Kojovic, Natraj, Mohanty, Maillart, and Schaer}]{kojovic2021using}
Nada Kojovic, Shreyasvi Natraj, Sharada~Prasanna Mohanty, Thomas Maillart, and Marie Schaer. 2021.
\newblock Using 2d video-based pose estimation for automated prediction of autism spectrum disorders in young children.
\newblock \emph{Scientific Reports}, 11(1):15069.

\bibitem[{Kuijper et~al.(2015)Kuijper, Hartman, and Hendriks}]{kuijper_who_2015}
Sanne J.~M. Kuijper, Catharina~A. Hartman, and Petra Hendriks. 2015.
\newblock Who {Is} {He}? {Children} with {ASD} and {ADHD} {Take} the {Listener} into {Account} in {Their} {Production} of {Ambiguous} {Pronouns}.
\newblock \emph{PloS One}, 10(7):e0132408.

\bibitem[{Lee and Lee(2023)}]{lee-lee-2023-lftk}
Bruce~W. Lee and Jason Lee. 2023.
\newblock \href {https://aclanthology.org/2023.bea-1.1} {{LFTK}: Handcrafted features in computational linguistics}.
\newblock In \emph{Proceedings of the 18th Workshop on Innovative Use of NLP for Building Educational Applications (BEA 2023)}, pages 1--19, Toronto, Canada. Association for Computational Linguistics.

\bibitem[{Lee et~al.(2012)Lee, Kim, Kim, Sung, and Kwon}]{cleft_palate}
Ji~Eun Lee, Wook~Eun Kim, Kwang~Hyun Kim, Myung~Whun Sung, and Tack~Kyun Kwon. 2012.
\newblock \href {https://doi.org/10.3342/kjorl-hns.2012.55.8.498} {Research on construction of the korean speech corpus in patient with velopharyngeal insufficiency.}
\newblock \emph{Korean J Otorhinolaryngol-Head Neck Surg}, 55(8):498--507.

\bibitem[{Lee et~al.(2023)Lee, Yeo, Kim, and Chung}]{lee_knowledge-driven_2023}
Seonwoo Lee, Eunjung Yeo, Sunhee Kim, and Minhwa Chung. 2023.
\newblock Knowledge-driven speech features for detection of korean-speaking children with autism spectrum disorder.
\newblock \emph{Phonetics and Speech Sciences}, 15:53--59.

\bibitem[{Lee et~al.(2015)Lee, Heo, and Jang}]{lssc}
YoonKyoung Lee, Hyunsook Heo, and Seungmin Jang. 2015.
\newblock \emph{Language Scale for School-aged Children ({LSSC})}.
\newblock Inpsyt.

\bibitem[{Li et~al.(2022)Li, Chen, Li, Ouyang, and Li}]{li2022automatic}
Jing Li, Zejin Chen, Gongfa Li, Gaoxiang Ouyang, and Xiaoli Li. 2022.
\newblock Automatic classification of asd children using appearance-based features from videos.
\newblock \emph{Neurocomputing}, 470:40--50.

\bibitem[{Lord et~al.(2012)Lord, Rutter, Luyster, and Gotham}]{lord_autism_2012}
Catherine Lord, Michael Rutter, Rhiannon~J. Luyster, and Katherine Gotham. 2012.
\newblock \emph{Autism diagnostic observation schedule-2nd edition ({ADOS}-2)}, 2nd edition edition.
\newblock Western Psychological Corporation.

\bibitem[{Lyakso et~al.(2017)Lyakso, Frolova, and Grigorev}]{lyakso_perception_2017}
Elena Lyakso, Olga Frolova, and Aleksey Grigorev. 2017.
\newblock Perception and acoustic features of speech of children with autism spectrum disorders.
\newblock In \emph{Speech and Computer}, Lecture Notes in Computer Science, pages 602--612. Springer International Publishing.

\bibitem[{{McCann} and Peppé(2003)}]{mccann_prosody_2003}
Joanne {McCann} and Sue Peppé. 2003.
\newblock Prosody in autism spectrum disorders: a critical review.
\newblock \emph{International Journal of Language \& Communication Disorders}, 38(4):325--350.

\bibitem[{McFee et~al.(2022)McFee, Metsai, McVicar, Balke, Thomé, Raffel, Zalkow, Malek, {Dana}, Lee, Nieto, Ellis, Mason, Battenberg, Seyfarth, Yamamoto, {viktorandreevichmorozov}, Choi, Moore, Bittner, Hidaka, Wei, {nullmightybofo}, Weiss, Hereñú, Stöter, Nickel, Friesch, Vollrath, and Kim}]{mcfee_librosalibrosa_2022}
Brian McFee, Alexandros Metsai, Matt McVicar, Stefan Balke, Carl Thomé, Colin Raffel, Frank Zalkow, Ayoub Malek, {Dana}, Kyungyun Lee, Oriol Nieto, Dan Ellis, Jack Mason, Eric Battenberg, Scott Seyfarth, Ryuichi Yamamoto, {viktorandreevichmorozov}, Keunwoo Choi, Josh Moore, Rachel Bittner, Shunsuke Hidaka, Ziyao Wei, {nullmightybofo}, Adam Weiss, Darío Hereñú, Fabian-Robert Stöter, Lorenz Nickel, Pius Friesch, Matt Vollrath, and Taewoon Kim. 2022.
\newblock \href {https://doi.org/10.5281/zenodo.6759664} {librosa/librosa: 0.9.2}.

\bibitem[{Pae and Kwak(2011)}]{k_mb_cid}
Soyeong Pae and Keum-Joo Kwak. 2011.
\newblock \emph{Korean MacArthur-Bates Communicative Development Inventories ({K M-B CDI})}.
\newblock Mindpress.

\bibitem[{Qualls and Corbett(2017)}]{socialcommunication}
Lydia~R Qualls and Blythe~A Corbett. 2017.
\newblock Examining the relationship between social communication on the ados and real-world reciprocal social communication in children with asd.
\newblock \emph{Research in autism spectrum disorders}, 33:1--9.

\bibitem[{Rapin et~al.(2009)Rapin, Dunn, Allen, Stevens, and Fein}]{rapin_subtypes_2009}
Isabelle Rapin, Michelle~A. Dunn, Doris~A. Allen, Michael~C. Stevens, and Deborah Fein. 2009.
\newblock Subtypes of language disorders in school-age children with autism.
\newblock \emph{Developmental Neuropsychology}, 34(1):66--84.

\bibitem[{Shriberg et~al.(2011)Shriberg, Paul, Black, and van Santen}]{shriberg_hypothesis_2011}
Lawrence~D. Shriberg, Rhea Paul, Lois~M. Black, and Jan~P. van Santen. 2011.
\newblock The hypothesis of apraxia of speech in children with autism spectrum disorder.
\newblock \emph{Journal of Autism and Developmental Disorders}, 41(4):405--426.

\bibitem[{Terpilowski(2019)}]{terpilowski_scikit-posthocs_2019}
Maksim~A. Terpilowski. 2019.
\newblock scikit-posthocs: Pairwise multiple comparison tests in python.
\newblock \emph{Journal of Open Source Software}, 4(36):1169.

\bibitem[{Turner et~al.(2006)Turner, Stone, Pozdol, and Coonrod}]{turner_follow-up_2006}
Lauren~M. Turner, Wendy~L. Stone, Stacie~L. Pozdol, and Elaine~E. Coonrod. 2006.
\newblock Follow-up of children with autism spectrum disorders from age 2 to age 9.
\newblock \emph{Autism: The International Journal of Research and Practice}, 10(3):243--265.

\bibitem[{Virtanen et~al.(2020)Virtanen, Gommers, Oliphant, Haberland, Reddy, Cournapeau, Burovski, Peterson, Weckesser, Bright, van~der Walt, Brett, Wilson, Millman, Mayorov, Nelson, Jones, Kern, Larson, Carey, Polat, Feng, Moore, VanderPlas, Laxalde, Perktold, Cimrman, Henriksen, Quintero, Harris, Archibald, Ribeiro, Pedregosa, van Mulbregt, and {SciPy 1.0 Contributors}}]{virtanen_scipy_2020}
Pauli Virtanen, Ralf Gommers, Travis~E. Oliphant, Matt Haberland, Tyler Reddy, David Cournapeau, Evgeni Burovski, Pearu Peterson, Warren Weckesser, Jonathan Bright, Stéfan~J. van~der Walt, Matthew Brett, Joshua Wilson, K.~Jarrod Millman, Nikolay Mayorov, Andrew R.~J. Nelson, Eric Jones, Robert Kern, Eric Larson, C~J Carey, Ilhan Polat, Yu~Feng, Eric~W. Moore, Jake VanderPlas, Denis Laxalde, Josef Perktold, Robert Cimrman, Ian Henriksen, E.~A. Quintero, Charles~R. Harris, Anne~M. Archibald, Antônio~H. Ribeiro, Fabian Pedregosa, Paul van Mulbregt, and {SciPy 1.0 Contributors}. 2020.
\newblock {SciPy} 1.0: Fundamental algorithms for scientific computing in python.
\newblock \emph{Nature Methods}, 17:261--272.

\bibitem[{Wetherby(2006)}]{wetherby2006understanding}
Amy~M Wetherby. 2006.
\newblock Understanding and measuring social communication in children with autism spectrum disorders.
\newblock \emph{Social and communication development in autism spectrum disorders: Early identification, diagnosis, and intervention}, 18(3):3--34.

\bibitem[{Wolk and Brennan(2013)}]{wolk_phonological_2013}
Lesley Wolk and Christine Brennan. 2013.
\newblock Phonological investigation of speech sound errors in children with autism spectrum disorders.
\newblock \emph{Speech, Language and Hearing}, 16(4):239--246.

\bibitem[{Wu et~al.(2023)Wu, Deng, Jian, Chen, Li, Gong, and Wu}]{dtx}
Xuesen Wu, Haiyin Deng, Shiyun Jian, Huian Chen, Qing Li, Ruiyu Gong, and Jingsong Wu. 2023.
\newblock Global trends and hotspots in the digital therapeutics of autism spectrum disorders: a bibliometric analysis from 2002 to 2022.
\newblock \emph{Frontiers in Psychiatry}, 14:1126404.

\end{thebibliography}


\end{document}